\documentclass{aa}  

\usepackage{graphicx}
\usepackage[colorlinks, linkcolor=blue, anchorcolor=blue, citecolor=blue]{hyperref}
\usepackage{txfonts}
\usepackage{color}
\usepackage{siunitx}
\usepackage{mathtools}
\usepackage{amsmath}
\usepackage{CJK}

\usepackage{natbib}
\bibpunct{(}{)}{;}{a}{}{,} 

\newcommand{\water}{{\rm H_2O}}

\newcommand{\frag}{{\rm frag}}

\newcommand{\dust}{{\rm dust}}
\newcommand{\surf}{{\rm surf}}

\begin{document}

\title{White dwarf magnetospheres: Shielding volatile content of icy objects and implications for volatile pollution scarcity}

\author{Wen-Han Zhou
       \inst{1}
       \and
       Shang-Fei Liu
       \inst{2,3}
       \and
       Douglas N.C. Lin
       \inst{4,5}
       }

\institute{Universit\'e C\^ote d'Azur, Observatoire de la C\^ote d'Azur, CNRS, Laboratoire Lagrange, Nice, France\\
           \email{wenhan.zhou@oca.eu}
     \and
     School of Physics and Astronomy, Sun Yat-sen University, Zhuhai, Guangdong Province, China\\
     \email{liushangfei@mail.sysu.edu.cn}
     \and
     CSST Science Center for the Guangdong-Hong Kong-Macau Greater Bay Area, Sun Yat-sen University, Zhuhai, Guangdong Province, China 
     \and
     Department of Astronomy and Astrophysics, University of California, Santa Cruz, USA
     \and
     Institute for Advanced Studies, Tsinghua University, Beijing, China}

\authorrunning{Zhou et al}

 
  \abstract
  {About 25\% -- 50\% of white dwarfs are found to be contaminated by heavy elements, which are believed to originate from external sources such as planetary materials. Elemental abundances suggest that most of the pollutants are rocky objects and only a small fraction of white dwarfs {{bear}} traces of volatile accretion.  
   }
   {In order to account for the scarcity of volatile pollution, we investigate the role of the white dwarfs' magnetospheres in shielding the volatile content of icy objects.}
   {
   We {estimated} the volatile sublimation of inward drifting exocomets. We assume the orbits of the exocomets are circularized by the Alfv{\'e}n wing drag that is effective for long-period comets. 
  
   }
   {
   Volatile material can sublimate outside the corotation radius and be shielded by the magnetic field. {The two conditions for this volatile-shielded mechanism are that the magnetosphere radius must be larger than the corotation radius and that the volatiles are depleted outside the corotation radius, which requires a sufficiently slow orbital circularization process.} We applied our model to nine white dwarfs with known rotational periods, magnetic fields, and atmosphere compositions. Our volatile-shielded model may explain the excess of volatile elements such as C and S in the disk relative to the white dwarf atmosphere in WD2326+049 (G29-38). Nevertheless, given the sensitivity of our model to the circularization process and material properties of icy objects, there remains considerable uncertainty in our results.
  }
   {We emphasize the importance of white dwarfs' magnetic fields in preventing the accretion of volatile gas onto them. Our work suggests a possible explanation for the scarcity of volatile-accretion signatures among white dwarfs. We also identify a correlation between the magnetic field strength, the spin period, and the composition of pollutants in white dwarf atmospheres. However, given the uncertainties in our model, more observations are necessary to establish more precise constraints on the relevant parameters.}


   \maketitle
%

\section{Introduction}

Between 25\% and 50\% of discovered white dwarfs have been found to contain heavy elements in their atmospheres \citep{Zuckerman2003, Koester2014}. The sinking timescale of these heavy elements is relatively short, ranging from several days (for hydrogen-dominated white dwarfs) to millions of years (for helium-dominated white dwarfs) \citep{Paquette1986a,Paquette1986b,Koester2009}. Therefore, these heavy elements are believed to originate from external sources other than the white dwarf core, due to the rapid sedimentation of heavy elements in white dwarfs \citep{Jura2003, Jura2008, Koester2006}. The hypothesis of interstellar origin of the pollution is ruled out by the inconsistency of the predicted element abundance with observation \citep{Dufour2007} and by the absence of correlation between the pollution and the space motions of white dwarfs \citep{Zuckerman2003,Koester2006}. It is widely believed that the pollution results from the accretion of planetary material, with supportive evidence from observed dusty disks and transiting planets around white dwarfs \citep{Koester2014,Rocchetto2015}. Therefore, the composition of pollutants in white dwarfs' atmospheres provides valuable information about the history of exoplanetary systems \citep{Harrison2018}. The relative abundances of various elements are indicative of the compositions of the pollutants. A total of 23 different elements have been detected in the polluted white dwarf atmospheres, including Ca, Si, Ma, Fe, and O, through optical spectroscopy, and C, N, and S, mainly through ultraviolet spectroscopy \citep{Zuckerman2007,Melis2016,Xu2021}. 

\begin{figure*}
    \centering
    \includegraphics[width = 1\textwidth]{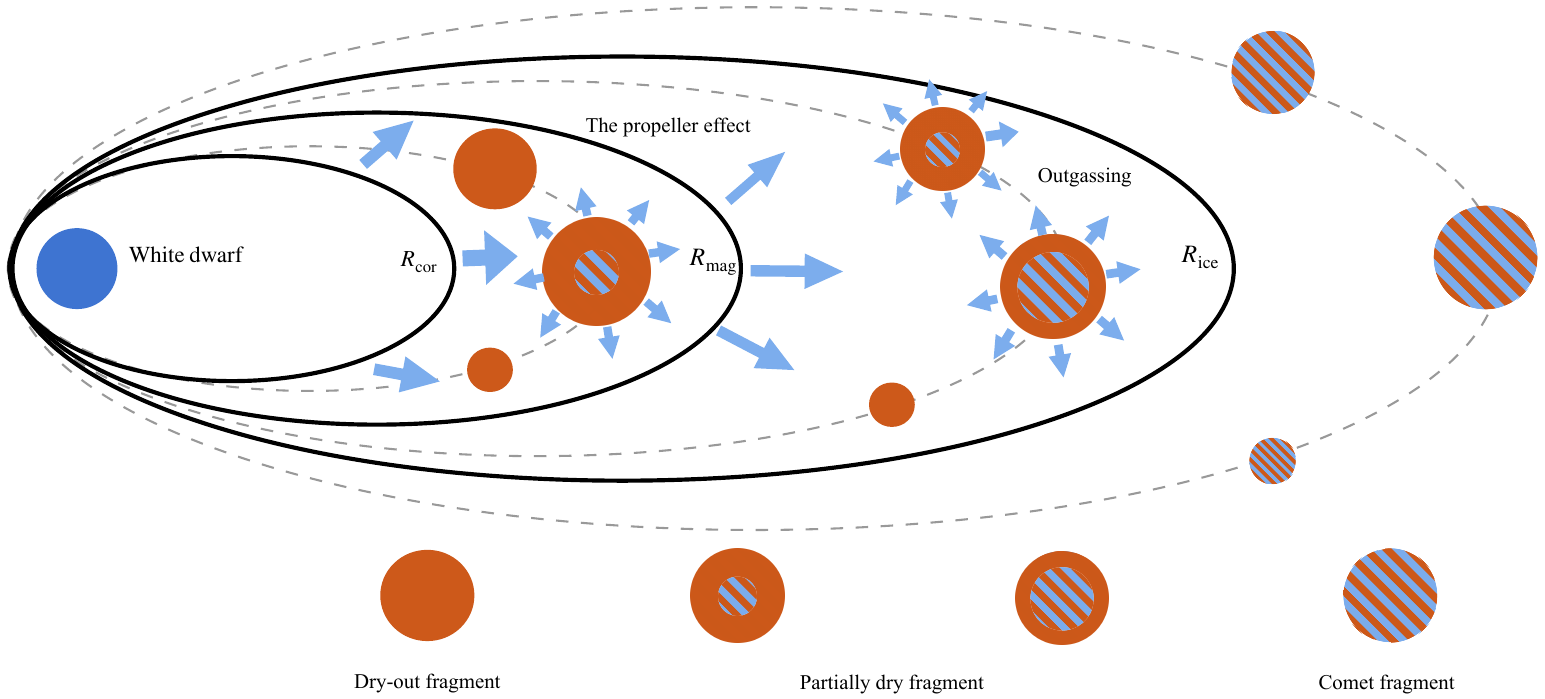}
    \caption{Schematic overview of volatile shielding by magnetic field. Three critical {{radii}} are denoted by black lines: the corotation radius $R_{\rm cor}$, the magnetosphere radius $R_{\rm mag}$, and the ice line radius $R_{\rm ice}$. The gradually circularized orbits of tidal fragments of a comet are denoted by gray {{dashed}} lines. In the process of the orbital circularization due to {{the magnetic drag}} near the pericenter, icy fragments gradually lose their water content by evaporation after crossing the ice line. Small fragments are dried out earlier than large fragments. Provided $R_{\rm mag} > R_{\rm cor}$, the vapor gas between $R_{\rm cor}$ and $R_{\rm mag}$ is propelled to infinity by the magnetic field, leading to a truncation of gas near $R_{\rm mag}$. Thus, the vapor gas is shielded by this propeller effect. If fragments vaporize all of their water content outside the corotation $R_{\rm cor}$, then only refractory materials cross the magnetosphere and finally get accreted onto the white dwarf. In this case   volatile accretion does not occur.}
    \label{fig:schematic_overview}
\end{figure*}

Interestingly, the dominant elements found in accreted material on white dwarfs are rock-forming elements such as Mg, Ca, and Fe, with only a few white dwarfs being polluted by icy materials. Volatile elements, such as C, N, and S, and excess oxygen (likely from water) are present in only a small number of polluted white dwarf atmospheres \citep{Farihi2013,Raddi2015,Xu2017}. Hydrogen, which does not sink in helium-dominated white dwarfs, can function as a record of water accretion in the white dwarf's history \citep{Veras2014, Gentile2017, Izquierdo2018}. Possible sources of volatile pollution in white dwarf atmospheres include hot gas giants resulting from exo-Kuiper belt or exo-Oort Cloud objects \citep{Jura2010,Malamud2016, Xu2017,Gansicke2019, Zhang2021}. The number of helium-dominated white dwarfs with detected hydrogen increases with the cooling age of the white dwarf, suggesting that water-rich bodies serve as a constant source of pollution over the long term. However, much less volatile elements are observed in white dwarf atmospheres compared to refractory elements \citep{Jura2011, Jura2014}, although the volatile materials in the interior of minor planets far away from the star  (i.e., tens of au) can survive from the intense luminosity during the red giant evolution \citep{Jura2010, Malamud2016, Malamud2017a, Malamud2017b}. {Several mechanisms} have been proposed to account for the scarcity of volatiles, including the primordially dry nature of the pollutants \citep{Harrison2021} and observational bias due to asynchronous accretion \citep{Malamud2016, Brouwers2022b}. However, recent research shows that volatile vapor is necessary to the accretion disk to offer a drag to the debris, thereby facilitating the high accretion rates deduced from observation \citep{Okuya2023}. This implies that volatiles could exist with refractory {elements} in the disk, but be prevented from accretion onto the white dwarf by some unknown physical process.

We propose a mechanism to explain the lack of volatile pollution, which involves the shielding of volatiles by the magnetic field in the scenario of comet accretion. A significant fraction (8\%--20\%) of white dwarfs have a magnetic field with strengths {{$0.1$ to $10^5$~T} }\citep{Ferrario2005, Ferrario2015}. 
 The magnetosphere radius $R_{\rm mag}$ is the distance at which the pressure from the magnetic field balances the ram pressure of the accreting gas. Inside $R_{\rm mag}$, the magnetic field significantly affects the orbital evolution of ionized gas, creating a funnel flow onto or propelling gas away from the white dwarf \citep{Pringle1972}, depending on the position of the corotation radius $R_{\rm cor}$. Ionized gas outside $R_{\rm cor}$ gains positive angular momentum in the magnetic field and moves outward from the white dwarf \citep{Illarionov1975, Lai2014}. Therefore, if $R_{\rm cor} < R_{\rm mag}$, the gas supply from outside $R_{\rm cor}$ is truncated at the corotation radius $R_{\rm cor}$ by the pressure from the magnetic field. When an icy object sublimates its volatile content before it crosses $R_{\rm cor}$, the vapor will be shielded by the propeller effect of the magnetic field and will not be accreted. We tested this hypothesis by investigating the thermal evolution and water evaporation of an icy object during its orbital circularization process.

 In the following, sublimation of the water ice in the drifting icy fragments is studied in Section \ref{sec2}, based on which we discuss {{the}} shielding effect of the magnetic field and the accreted mass fraction of water content in the icy objects under different conditions in Section \ref{sec3}. In Section \ref{sec4} we apply our model to a few white dwarfs with known magnetic fields and rotation periods.  
 \section{Evolution of a tidally fragmented comet}
\label{sec2}

We consider the typical lifespan of an exocomet that ultimately accretes onto a white dwarf. The comet becomes tidally disrupted upon being scattered into the Roche radius, causing fragments to spread onto highly eccentric orbits. These fragments subsequently undergo orbital circularization due to Alfvén-wave drag by the magnetic field {{\citep{Zhang2021}}} or dust drag by {{a}} pre-existing disk {{\citep{Malamud2021}}}. Once the fragments cross the ice line, the volatiles inside vaporize and release from the surfaces. Thus, the thermal evolution of the comet fragments is closely linked to their orbital evolution.

\subsection{Orbital {circularization}}
\label{sec:orbital evolution}
A comet breaks up due to tidal force once it is within the tidal disruption radius $R_{\rm tidal}$. For a comet larger than {{200~m}}, its material strength is gravity-dominated. In this case, the tidal disruption radius is 
\begin{equation}
    R_{\rm tid,grav} = 1.05 \left( \frac{M_\star}{\rho} \right)^{1/3} = 1.27 \left(\frac{\rho}{2\rm~g/cm^3} \right)^{-1/3} \left(\frac{M_\star}{0.6 M_\odot} \right)^{1/3} R_\odot.
\end{equation}
The size distribution of fragments is well described by a power law with a power index of -3.5, which results from a collisional grind-down of fragments. The maximum size of fragments is {{assumed to be $\sim 200~$m, since larger fragments are considered to be cohesionless and are further tidally fragmented, based on the material properties of small bodies in the Solar System \citep{Walsh2018}}}. Fragments smaller than a few microns are blown out by radiation pressure {{\citep{Murray1999, Brouwers2022a}}}. These fragments are distributed on highly eccentric orbits, and depending on the initial orbit of the comet, some may even be unbound. Further analysis of the size distribution and orbital distribution of fragments can be found in Appendix \ref{AppA}.

The orbits of fragments become circularized due to various drags, such as the {Poynting–Robertson (P-R)} drag for small fragments (<1~cm), and the dust drag \citep{Jura2008,Veras2015, Malamud2021, Brouwers2022a} and Alfv{\'e}n-wave drag \citep{Zhang2021} for larger fragments. 

{{When conductive objects move through a magnetized flowing plasma  (e.g., ionized gas particles present within the refractory or volatile sublimation line of the white dwarf), a loop is formed by the induced Alfv{\'e}n wave connecting the fragment to the surrounding plasma \citep{Drell65, Bromley2019}. The object acts as a battery and dissipates its orbital energy. This phenomenon has been extensively explored in various astrophysical contexts, such as asteroids or comets orbiting a young Sun or a T Tauri star \citep{Sonett1970}, the heating of asteroids around a white dwarf \citep{Bromley2019} or a pulsar \citep{Kotera2016}, as well as the unipolar induction observed in the moons of Jupiter and Saturn \citep{Hand2011}. }} The timescale of this orbital circularization is 
\begin{align}
    \tau_{B,circ} &= {2\pi \sqrt{GM_\star a_0 } \over -\Delta E}    \\
    & = 2.7 \left(\frac{r}{1~\rm m} \right)\left(\frac{B_\star}{100~\rm T} \right)^{-2} \left(\frac{a_0}{100~\rm au} \right)^{1/2}  \left(\frac{q_0}{0.2 R_\odot} \right)^{11/2}  {~\rm kyr,}
\end{align}
where $a_0$ is the initial semimajor axis before orbital circularization, $R_{\rm p}$ is the pericenter of the fragments, and $B_\star$ is the magnetic field on the white dwarf surface. The detailed derivation of this timescale is given in Appendix \ref{AppB}. 

Some white dwarfs are discovered to hold a dust disk \citep{Rocchetto2015}, a small fraction of which are also gaseous \citep{Manser2020}. If these white dwarfs do not have a strong magnetic field, the drag caused by a pre-existing dusty disk could dominate the orbital circularization of fragments. The timescale is given by \citep{Malamud2021}
\begin{equation}
\begin{aligned}
    \tau_{\rm D, circ} &= \frac{4\pi \rho r p}{3C_{\rm d} \Sigma_{\rm disk}} \sqrt{\frac{a_0}{G M_\star}}  \\
    & \sim 0.45 \left(\frac{r}{1~\rm m} \right) \left(\frac{a_0}{100~\rm au} \right)^{1/2} \left(\frac{q_0}{0.2 R_\odot} \right)^{5/2} \left(\frac{M_{\rm disk}}{10^{19}~\rm g} \right)^{-1}    \rm kyr,
\end{aligned}
\end{equation}
where $C_{\rm d}$ is of order unity and $\Sigma_{\rm disk}$ is the surface density of the disk that gives the disk mass $M_{\rm disk}$. 

Assuming the presence of a compact disk, the timescale for circularization caused by dust drag is shorter than that caused by Alfvén-wave drag. However, the existence of a pre-existing disk is not always observed as only a small number of white dwarf systems host dust disks \citep{Farihi2016}. Moreover, this mechanism may not be effective for long-period comets, which are the focus of study in this paper, as the lifetime of the compact disk could be shorter than the orbital periods of fragments \citep{Malamud2021}. Therefore, in this paper, we mainly employ Alfvén-wave drag-induced circularization to investigate the orbital evolution of fragments, although dust drag-induced circularization may be applicable in some cases.

\subsection{Water evolution}
\label{sec:analytical_model}
Fragments experience both orbital decay and circularization due to the Alfv{\'e}n-wave drag, leading to an increase in mean temperature over one orbital period, and consequent heating of the fragment's interior. Once fragments cross the ice line, volatiles in the interior of comets are released through the porous structure as a result of sublimation. Considering the timescale considered in this work is long ({on the order of a million years}), we used an approximate analytical model to save computation time.

{Comets consist of a complex mixture of ices, primarily water and carbon dioxide, alongside dust predominantly made of silicates and carbon-based materials \citep{Huebner2006}. Water makes up of $\sim 75\%$ volatiles with the rest dominated by carbon dioxide and carbon monoxide \citep{Fulle2019, Davidsson2021}. This paper focuses primarily on the evolution of water ice for the sake of simplicity. Incorporating additional volatiles into our analysis would necessitate a more sophisticated numerical model, which falls beyond the scope of our current study. Notably, volatiles such as CO and $\rm CO_2$ have significantly lower sublimation temperatures compared to water, enabling them to sublimate at greater distances from the star. This characteristic allows them to be more readily shielded by the magnetic field.}

\begin{figure}
    \centering
    \includegraphics[width=0.5\textwidth]{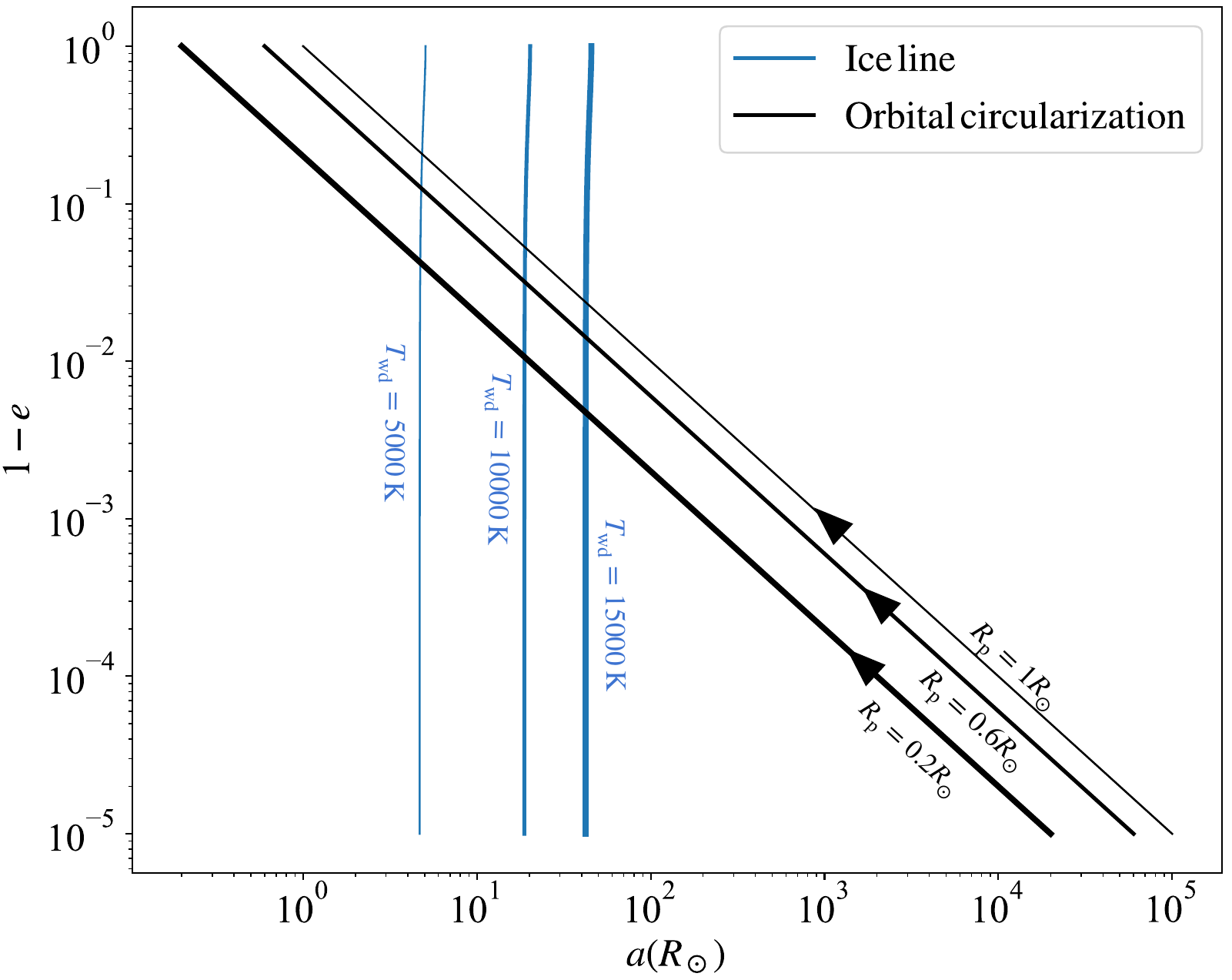}
    \caption{Ice lines for water in the white dwarf system, denoted by blue lines with different line widths with respect to different white dwarf temperatures. The black lines show the process of the orbital circularization with arrows representing the evolving direction and different line widths for different pericenters.}
    \label{fig:ice_line}
\end{figure}

{Based on observations of cometary activity and in situ measurements on their surfaces, comets are often described as dusty snowballs enveloped by a dry dust mantle \citep{Whipple1950, Hu2017, Pajola2017}. The depletion of volatiles from these icy bodies can be estimated through an energy balance approach. Considering a spherical icy fragment with a radius denoted as $r_\frag$,  the evolution of the water mass in an icy fragment occurs at a rate of }
\begin{equation}
\label{eq:m_dot_H2O}
    \dot m_{\rm H_2O} = - 4 \pi r_\water^2 J_\water. 
\end{equation}
{Here $r_\water$ is the radius of the water ice interface at a given time $t$. The sublimation flux $J_\water$ can be estimated by simply assuming the thermal flux conducted to the ice layer is entirely absorbed by sublimation of water ice:}
\begin{equation}
\label{eq:sublimation_flux}
    J_\water L_\water = k_\dust {T_\surf - T_\water \over r_\frag - r_\water}.
\end{equation}
{Here $L_\water = 2.8~$MJ/kg is the latent heat of the water. The thermal conductivity of the dry dust layer $k_\dust$ is roughly 0.1~W/m/K, ignoring its dependence on the temperature  \citep[e.g., Table 2.6 in][]{Huebner2006}. The sublimation temperature of water ice $T_\water$ is around 180~K, and the surface temperature $T_\surf$ of an object on an eccentric orbit can be estimated as \citep{Gkotsinas2022}}
\begin{equation}
\label{eq:T_surf}
    T_\surf = \left( \frac{L_{\rm wd}}{16\pi \sigma a^2 (1 +  e^2/8 + 21e^4/512)^2 }  \right)^{1/4}
,\end{equation}
{with $L_{\rm wd} \simeq 0.001 L_\odot$ denoting the luminosity of the white dwarf. By setting $T_{\rm sur} = 180$~K, we obtain the iceline of the white dwarf, which gives us a quick idea of where the water starts to sublimate significantly, as shown in Fig.~\ref{fig:ice_line}.}

{It is important to note that the semimajor axis $a$ and eccentricity $e$ change over time, due to orbital circularization caused by the magnetic drag (see Eq.~\ref{eq:a}). Therefore, to model the evolution of the water fraction under the influence of orbital circularization, we follow these steps:}
\begin{enumerate}
    \item {We initiate the model with the icy object placed at $a=$400 $R_\odot$, well beyond water ice line. The pericenter $R_{\rm p}$ is chosen as a free parameter that ranges from 0.2 $R_\odot$ to $R_\odot$. The initial depth of the water-ice interface is set as 0.2~m, which is comparable to or slightly larger than the thermal skin.}
\item {The rate of water loss is calculated using Eqs.~\ref{eq:m_dot_H2O}, \ref{eq:sublimation_flux}, and \ref{eq:T_surf}.}
\item {The water mass $m_\water$ is updated for a time increment $\Delta t$ by adding $\dot m_\water \Delta t$ to the exsiting $m_\water$. The time step $\Delta t$ is chosen to be $10^{-3}$ of the circularization timescale, $\tau_{B, \rm circ}$, as defined in Eq.~\ref{eq:tau_B,circ}. Subsequently, the radius of the water ice $r_\water$ is updated based on $m_\water$.}
\item {Steps 2 and 3 are repeated until the orbit has fully circularized.}
\end{enumerate}
{As the outgassing process of water progresses, the surface of the icy object is eroded since small dust grains are carried out of the surface and become unbound from the object. The dust-to-gas mass ratio in cometary comae shows a wide range from 1 to 10, as estimated from different distances and measurements \citep{Rotundi2015, Fulle2016, Choukroun2020}. In this work we assume that the dust-to-gas mass ratio in the ejecta is 1. The total refractory-to-volatile mass ratio in an icy object is assumed to be 4, based on findings from the Rosetta mission to comet 67P/Churyumov-Gerasimenko \citep{Fulle2017, Fulle2019}.  Additionally, considering a density ratio of 2 between the dust and water ice, the dry layer of the fragment is eroded at a rate equivalent to 1/8 of the water retreat rate. To simulate this erosion process, we progressively reduce the dry layer based on the amount of water vapor released at each time step.} 

{This simple approach allows us to estimate the water loss during orbital circularization and to evaluate our hypothesis concerning the shielding effect of the magnetosphere. For the sake of simplicity, we focus solely on orbital circularization due to magnetic drag, although future studies could explore other circularization mechanisms, such as gas drag from a pre-existing disk, as well as the evaporation of additional volatiles in greater detail.}

\begin{figure}
    \centering
    \includegraphics[width=0.5\textwidth]{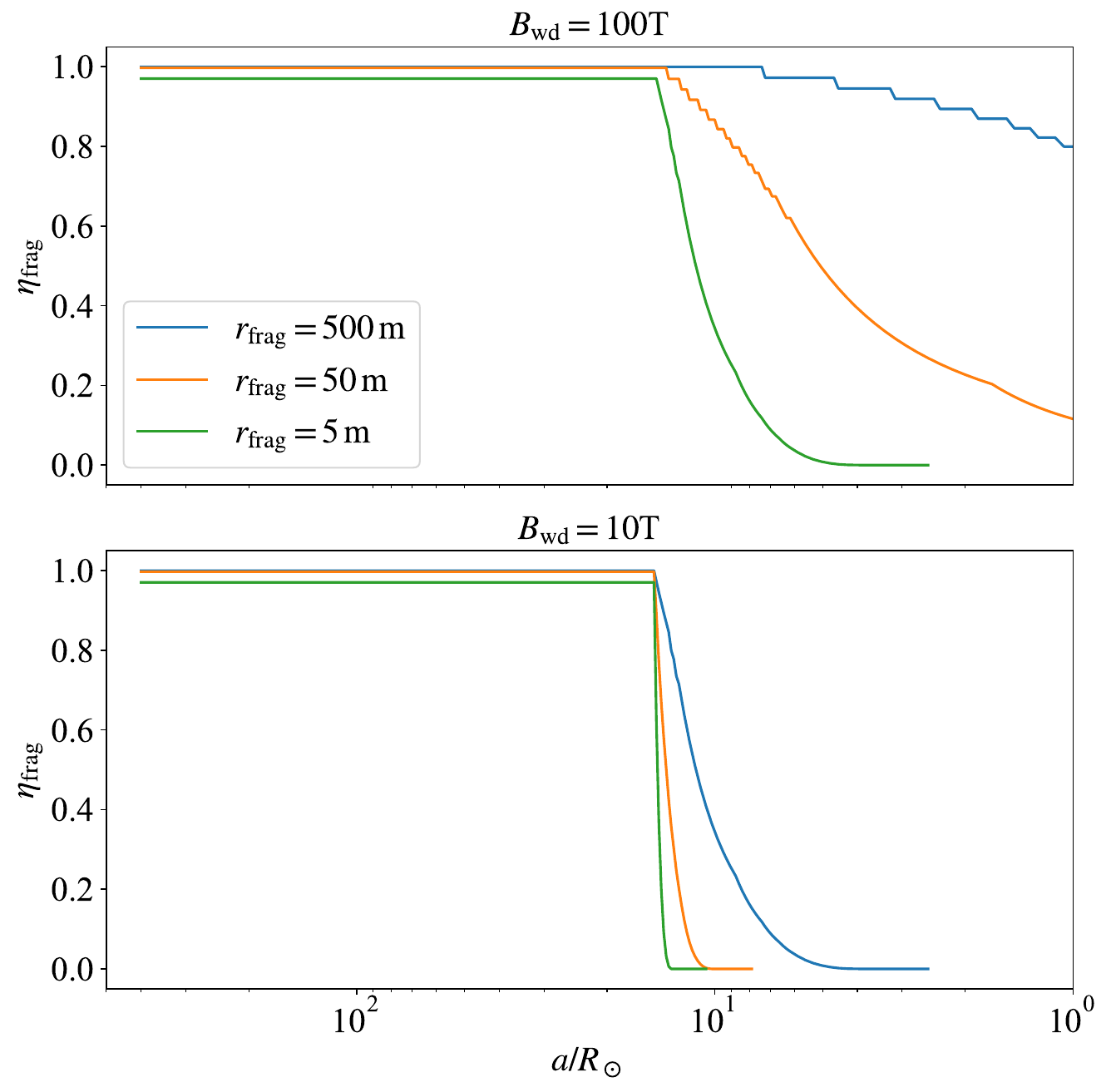}
    \caption{Residual mass fraction of water $\eta_{\rm frag}$ in drifting fragments of different sizes as a function of the semimajor axis. The corotation radius $R_{\rm cor} = 3.3R_\odot$ for a period of one day is denoted by the purple dashed line.}
    \label{fig:mass_evolution}
\end{figure}

\begin{figure}
    \centering
    \includegraphics[width=0.5\textwidth]{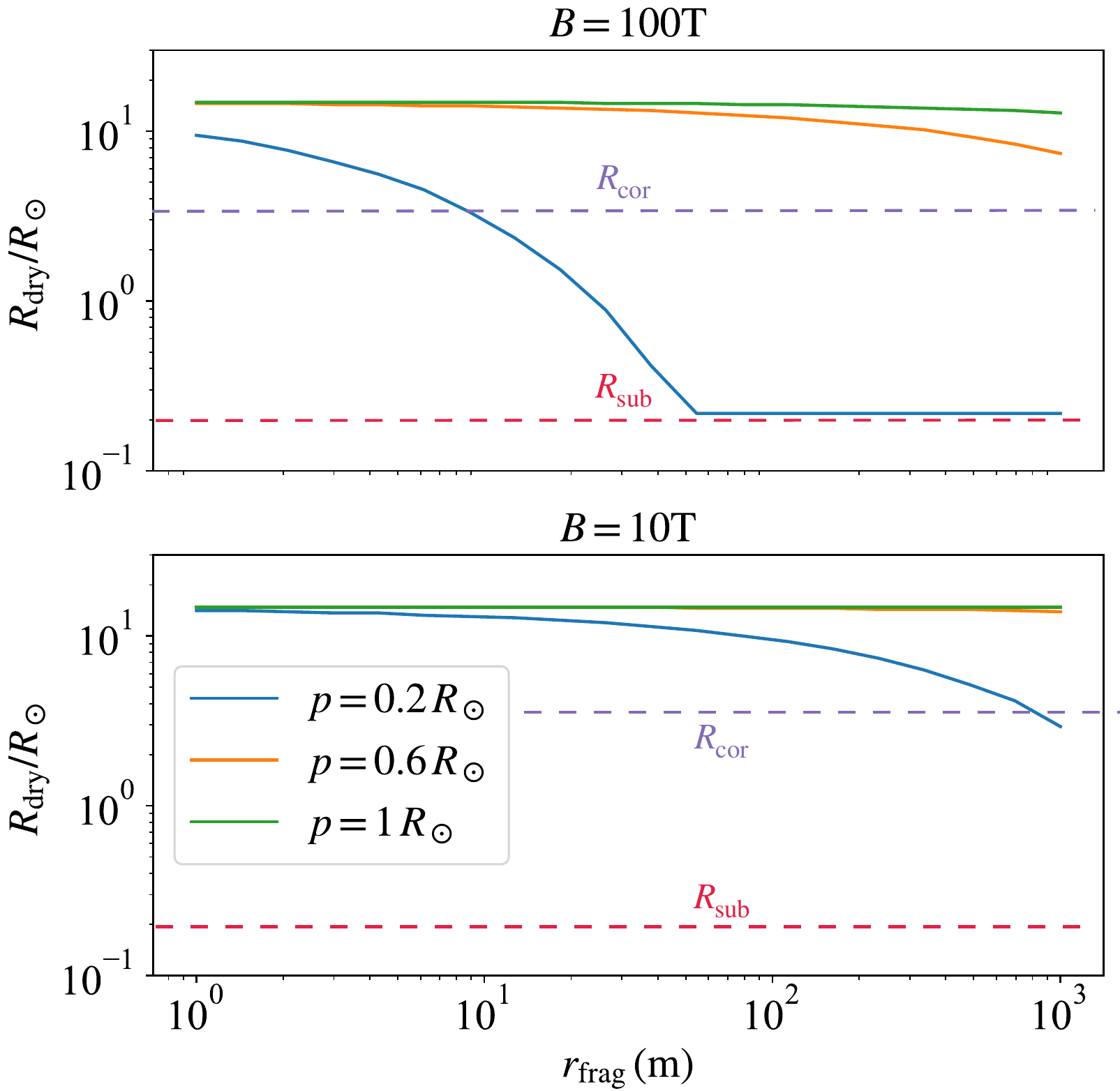}
    \caption{Dry-out distance $R_{\rm dry}$ for fragments as a function of the fragment size, accounting for different magnetic fields at the surface of the white dwarf. The corotation radius $R_{\rm cor} = 3.3R_\odot$ for a period of one day and the sublimation radius $R_{\rm sub} = 0.2 R_\odot$ for refractory sublimation is denoted by the purple dashed line and the red dashed line, respectively.}
    \label{fig:dry_radius}
\end{figure}

{Fragments may become completely dry at the orbital semimajor axis $a_{\rm dry}$ during the process of orbital circularization caused by the Alfv{\'e}n-wave drag. Figure \ref{fig:mass_evolution} illustrates the evolution of the mass fraction of residual volatile in the icy fragments. In a strong magnetic field ($B_{\rm} = 100~$T) the large fragments may not dry out before reaching the corotation radius $R_{\rm cor}$ as the fast circularization shortens the evaporation time. In contrast, in a weaker magnetic field ($B_{\rm} = 10~$T) fragments as large as 500~m can be completely dried out. To account for the orbital evolution of fragments with different pericenters, Figure \ref{fig:dry_radius} shows the dry-out radius $R_{\rm dry}$. As the pericenter $R_{\rm p}$ increases, fragments feel weaker magnetic fields, leading to a longer evaporation time and a more thorough evaporation.}

\section{Volatile-shielding by the magnetic field }
\label{sec3}

The evaporated vapor is rapidly photoionized on a timescale significantly shorter than the timescale of orbital {circularization}, as demonstrated in Appendix \ref{AppC}. Concurrently, the recombination rate is notably slower, resulting in near-complete ionization of the gas in the photoionization equilibrium, as explained in Appendix \ref{AppD}. The ionized gas interacts with the magnetic field and is either funneled onto the white dwarf or ejected to infinity \citep{Lai2014}.

\subsection{Magnetosphere radius}
Beyond the corotation radius $R_{\rm cor}$, the particle's motion is slower than that of the local magnetic field lines, resulting in a force that resists its relative movement. Consequently, the particle accumulates angular momentum and experiences a net outward force induced by the magnetic field, prompting it to move away from the white dwarf. On the other hand, within the corotation radius $R_{\rm cor}$, the particle loses both its energy and angular momentum and is subjected to a net inward force, directing it toward the white dwarf. The corotation radius is a function of the white dwarf angular speed $\omega_{\rm wd}$ 
\begin{equation}
\begin{aligned}
    R_{\rm cor} &= \left( \frac{G M_{\rm wd}}{\omega_{\rm wd}^2} \right)^{1/3} \\
    &= 3.3 R_{\odot} \left( \frac{M_{\rm wd}}{0.6 M_\odot} \right)^{1/3} \left( \frac{P_{wd}}{1~{\rm day}} \right)^{2/3},
\end{aligned}
\end{equation}
where $P_{wd} = 2\pi/\omega_{\rm wd}$ is the spin period of the white dwarf.

The critical radius within which the magnetic field dominates the incoming ionized gas flow is  called the magnetosphere radius, which can be obtained by
\begin{equation}
\label{eq:R_mag}
    R_{\rm mag} = R_{\rm wd} \left( \frac{B_\star^2}{\mu_0 \rho_g v_g^2} \right)^{1/6},
\end{equation}
where $\rho_g$ is the mass density of the gas. To ensure the magnetic effect, the magnetosphere radius $R_{\rm mag}$ must exceed the corotation radius $R_{\rm cor}$. In the case where the volatiles in the comet fragments sublimate beyond the corotation radius, they are ionized and shielded by the magnetic field. Nevertheless, for larger fragments, some volatiles may still remain in the deep interior as they cross the corotation radius.
The number density $\rho_g$ can be obtained by
\begin{equation}
\label{eq:rho_g}
    \rho_g = \frac{\dot M}{4 \pi R_{\rm mag}^2 v_g }
\end{equation}
as a first approximation. With Equations (\ref{eq:rho_g}) and (\ref{eq:v_g}) we find the analytical solution of $R_{\rm mag}$
for spherical accretion 
\begin{equation}
\begin{aligned}
    R_{\rm mag} &= R_{\rm wd} \left( \frac{16 \pi^2 B^4 R_{\rm wd}^5}{\mu_0^2 \dot M^2 G M_{\rm wd}} \right)^{1/7} \\
    &= 80 R_\odot \left( \frac{B_\star}{100\rm ~T} \right)^{4/7} \left( \frac{\dot M}{10^5 \rm ~kg/s} \right)^{-2/7}.
\end{aligned}
\end{equation}
For the magnetic shielding mechanism to work, the requirement of $R_{\rm mag} > R_{\rm cor}$ gives
\begin{equation}
\label{eq:condition1}
     \left(\frac{B_\star}{1\rm ~T} \right)^{4/7} \left( \frac{\dot M}{10^5 \rm ~kg/s} \right)^{-2/7} > \left( \frac{M_{\rm wd}}{0.6 M_\odot} \right)^{1/3} \left( \frac{P_{wd}}{2~{\rm day}} \right)^{2/3}.
\end{equation}

\begin{figure}
    \centering
    \includegraphics[width = 0.5\textwidth]{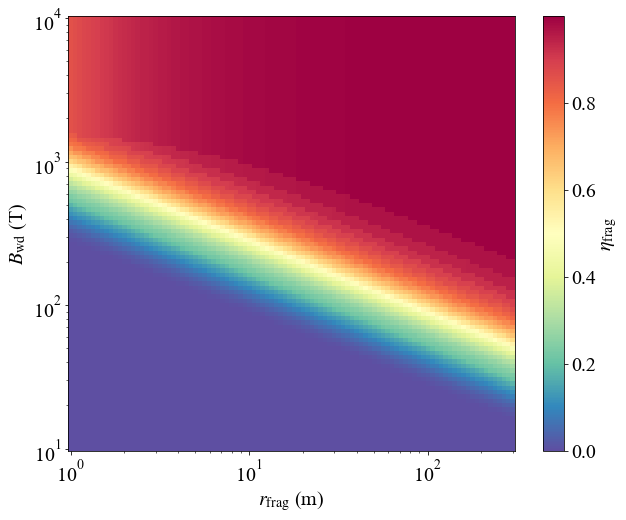}
    \caption{Residual mass fraction of water $\eta_{\rm frag}$ in fragments of different sizes under the different magnetic fields.}
    \label{fig:my_label}
\end{figure}


\subsection{Accreted mass fraction of the water in the comet}
\begin{figure}
    \centering
    \includegraphics[width = 0.5\textwidth]{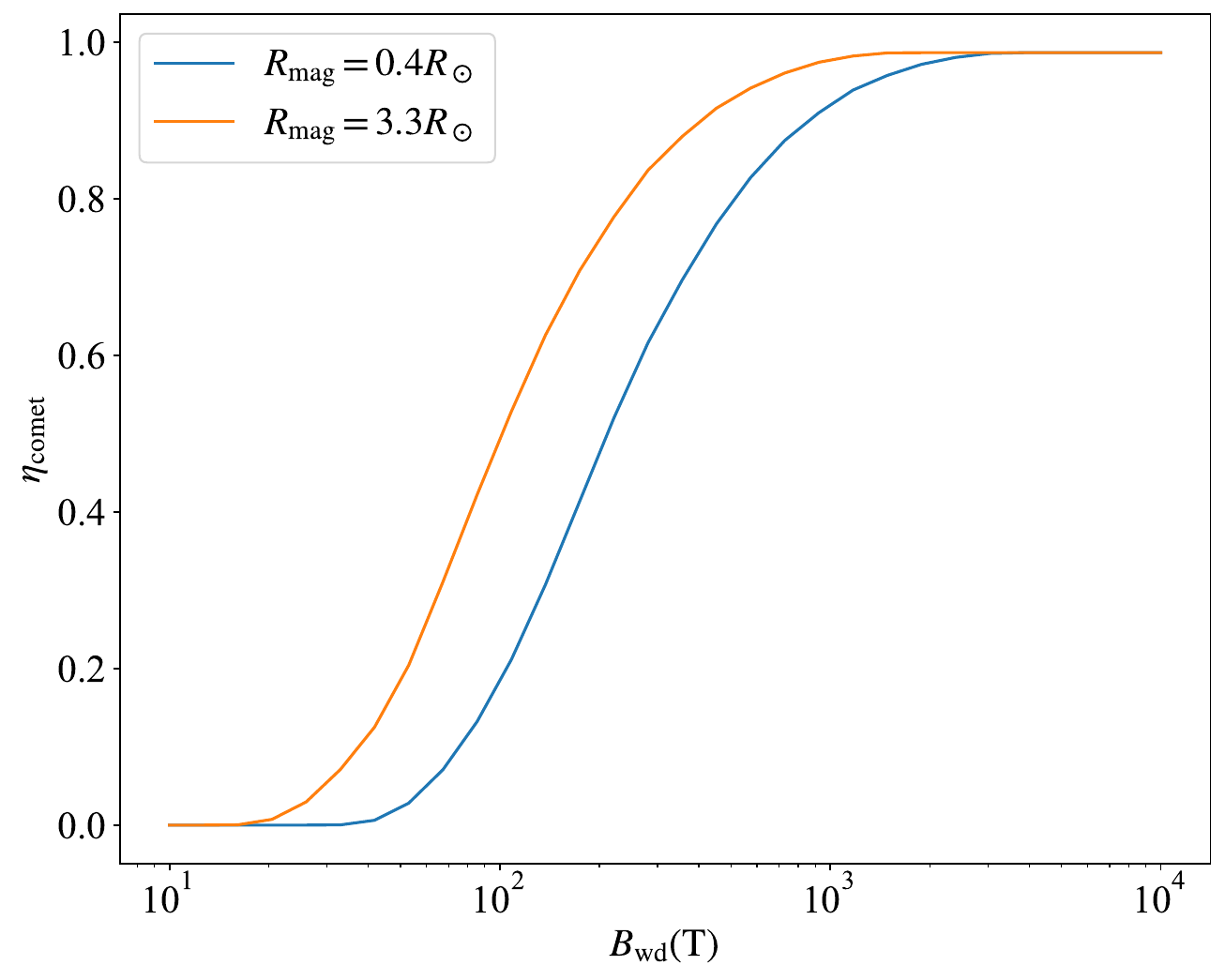}
    \caption{Accreted mass fraction of the volatile in the comet as a function of the magnetic field in the white dwarf surface, given $R_{\rm mag} \geq R_{\rm cor}$. {Two chosen values 0.4~$R_\odot$ and 3.3~$R_\odot$ correspond to the values of $R_{\rm cor}$ when the spin period of the white dwarf is one hour and one day, respectively.} }
    \label{fig:comet_mass_ratio}
\end{figure}

The tidal fragments of the comets follow a mass-size distribution 
\begin{equation}
    {\rm d} m = \frac{4\pi}{3}\rho(4- \alpha)r_0^3 r_{\rm frag,max}^{\alpha-4}r_{\rm frag}^{3-\alpha} {\rm d} r_{\rm frag},
\end{equation}
with which   the accreted mass of a comet can be obtained  by
\begin{equation}
    m_{acc} = \int_{=\infty}^{r_{\rm frag,max}} \eta_{\rm frag}(B,r_{\rm frag}) {\rm d}m.
\end{equation}
Therefore, the fraction of the volatile mass that could finally be accreted onto the white dwarf is 
\begin{equation}
    \eta_{\rm comet} = \frac{m_{acc}}{m_{comet}}.
\end{equation}
Figure \ref{fig:comet_mass_ratio} shows the residual mass fraction $\eta_{frag}(B,r_{\rm frag})$ of the volatile that is retained in the fragment after crossing the corotation radius $R_{\rm cor}$. 

\section{Discussion}
\label{sec4}

\subsection{Conditions for volatile shielding}
There are two critical requirements for efficient volatile shielding by the magnetic field. In the magnetic field, ionized gas particles outside the corotation radius $R_{\rm cor}$ obtain a positive orbital angular momentum and are consequently expelled from the white dwarf \citep{Lai2014}. To ensure the effectiveness of this shielding mechanism, the magnetic pressure must dominate over the gas ram pressure, constituting the first condition,
\begin{equation}
    R_{\rm mag} > R_{\rm cor},
\end{equation}
which is equivalent to the inequality (\ref{eq:condition1}). This magnetosphere condition gives the maximum rotational period of the white dwarf as a function of the magnetic field strength.

The second condition is referred to as the dry-out condition. Comets begin to sublimate water ice as they cross the ice line $R_{\rm ice}$ and continue until all volatile material is exhausted. In order for the magnetic field to effectively shield the volatile material, the corotation radius must lie within the ice line so that the volatile material can be sublimated into space before the comet enters the corotation radius. Therefore, we have
\begin{equation}
    R_{\rm ice} > R_{\rm cor},
\end{equation}
which gives the maximum rotational period of the white dwarf as a function of the temperature. Most of the volatiles need to be sublimated out of the comet's surface before the comet gets into the corotation radius. The {{semimajor axis}} where the icy object becomes completely dried out is called the dry-out radius $R_{\rm dry}$. In order for complete volatile shielding to occur, we require
\begin{equation}
    R_{\rm dry} > R_{\rm cor}
\end{equation}
or 
\begin{equation}
    \tau_{\rm dry} < \tau_{\rm circ}
\end{equation}
which means that the comet is dried out faster than its orbital decay. The dry-out radius of a comet is an important factor that determines whether volatile shielding by the magnetic field is effective. The dry-out radius is influenced by several physical properties of the comet, including its size, density, porosity, and thermal parameters, as well as the heating process it undergoes. While the properties of Solar System comets are used in this work, they are still uncertain. In general, larger comets have smaller dry-out radii because they require more time to sublimate their volatile content, leading to a closer approach to the white dwarf when completely dried out. The largest size of the tidal fragments, determined by material cohesion, can also influence the model results. The heating process is dependent on the temperature of the white dwarf, which defines the ice line radius, and the orbital circularization of the comet, which defines the dry-out radius. The conditions for volatile shielding by the magnetic field are illustrated in Figure \ref{fig:condition}, which considers different accretion rates, strengths, and periods of the white dwarf.

The model presented in this study involves a number of factors that introduce a significant degree of uncertainty to the results. The magnetosphere condition is primarily dependent on the physical properties of the white dwarf, while the dry-out condition is related to the thermal and orbital {circularization} of the icy object. Although this work focuses specifically on the water content of comets, the same analysis could be extended to other volatiles such as CO and $\rm CO_2$. It should be noted that the magnetic field may also shield rocky materials from the white dwarf, provided that the corotation radius is closer than the sublimation radius of such materials. For instance, the sublimation radius for silicate is estimated to be around $0.2 R_\odot$, which is also often regarded as the inner edge of a dusty disk \citep{Rafikov2011}. If $R_{\rm cor} < 0.2 R_\odot$, the accretion of planetary material would be significantly impeded. A corotation radius this close to the white dwarf is indicative of a fast-rotating object with a rotational period of less than 20 minutes.

For white dwarfs with strong magnetic fields, the magnetosphere condition ($R_{\rm mag} > R_{\rm cor}$) can generally be fulfilled except in the case of extremely slow rotators. However, satisfying the dry-out condition may be more difficult as icy fragments can still contain significant amounts of volatile material when they cross the corotation radius, due to rapid orbital circularization under the influence of the strong magnetic field. As shown in Figure \ref{fig:condition}, the dry-out condition becomes increasingly difficult to satisfy as the magnetic field strength of the white dwarf exceeds $10^4$~T.

In the case of weakly magnetic white dwarfs, the dry-out condition is satisfied due to the long orbital circularization time resulting from the weak magnetic field. As long as the magnetosphere condition is fulfilled, the shielding mechanism can effectively prevent volatiles from accreting onto the white dwarf. However, the spin period of the white dwarf should be less than one day for accretion rates ranging from $10^3~$kg/s to $10^7~$kg/s, as illustrated in Figure \ref{fig:condition}. 
 The accretion rate can be affected by the size of the pollutant. Accretion of exomoons could result in a high accretion rate, while consecutive asteroids produce a lower accretion rate \citep{Trierweiler2022}. Our results show that a higher accretion (e.g., from exomoons) {{is}} more likely to escape the volatile-shielding regime and cause volatile pollution as in this case the magnetosphere radius could be smaller than the corotation radius. It is important to note that Figure \ref{fig:condition} is based on the assumption that the Alfv{\'e}n drag dominates the orbital circularization process, which may not always be true. For instance, a weak magnetic field with a pre-existing compact dust disk could result in a quicker circularization process, thereby reducing the effectiveness of the volatile shielding mechanism, as discussed in Section \ref{sec3}. Additionally, for exo-Kuiper belt objects, the effect of dust drag may further impede the volatile shielding mechanism, while for exo-Oort Cloud objects its effect is minimal.


\subsection{Observation tests}
\begin{table*}[]
\caption{White dwarfs analyzed in this work.}
     \centering
     \begin{tabular}{l l l l l l l}
      \hline
      \noalign{\smallskip}
      {\rm Name} & {\rm Composition} & {\rm Surface magnetic field} & Temperature &{\rm Spin period} & {\rm Model (volatile)} & References \\
      \hline
      \noalign{\smallskip}
            0041-102 & DBA & 3500~T & 20000~K & 131.6~min & unshielded & 1,2,3 \\
            0853+163 & DBA & 100~T & 21200-27700~K& 2-24~hr &unshielded &4,5,6,7 \\
            1820+609 & DBA & <10~T & 4780~K& ~yrs &unshielded&5,6,8 \\
            1829+547 & DBA & 17000-18000~T & 6300~K & >100~yr &unshielded&8,9,10,11 \\
            0912+536 & DB & 10000~T & $7160 \pm 190$~K& 1.33~d & unshielded&1,8,12 \\
            1748+708 & DB & 10000~T & $5590 \pm 90$~K & >100~yr &unshielded &8,13,14,15 \\
            2010+310 & DB & 52000~T & 18000~K & >100 ~yr &unshielded &15,16,17 \\
            0322-019 & DAZ & $1.65 \pm 0.23$~T & $5310 \pm 100$~K & 28-33~d &unshielded &4, 18, 19 \\
            2326+049 & DAZ & $0.07 \pm 0.05$~T & 11900~K & 0.014~d; 0.042~d; 1.35~d &shielded &19,20,21,22 \\
      \noalign{\smallskip}
      \hline
     \end{tabular}
     \tablebib{(1)~\citet{Angel1977}; (2)~\citet{Achilleos1992}; (3)~\citet{Liebert1977}; (4)~\citet{Kawka2012}; (5)~\citet{Putney1997}; (6)~\citet{Brinkworth2013}; (7)~\citet{Wesemael2001}; (8)~\citet{Bergeron2001}; (9)~\citet{Putney1995}; (10)~\citet{Angel1975}; (11)~\citet{Silber1994}; (12)~\citet{Angel1972}; 
 (13)~\citet{Angel1974}; (14)~\citet{Angel1978}; (15)~\citet{Berdyugin1999}; (16)~\citet{Wickramasinghe2002}; (17)~\citet{Angel1985};
 (18)~\citet{Farihi2011}
 (19)~\citet{Farihi2018}
 (20)~\citet{Koester1998}
 (21)~\citet{Berger2005}
 (22)~\citet{Kleinman1998}
}
     \label{tab1}
 \end{table*}
Our model demonstrates a correlation between volatile accretion {{and}} the temperature, rotational {{periods}}, and magnetic field of the white dwarf. To our knowledge, only a limited number of white dwarfs have a known period, including $\sim 40$ observed through the non-radial $g$-mode pulsations of white dwarfs \citep[see Table 4 in][]{Hermes2017} and $\sim 35$ through photometry and spectroscopy of magnetic white dwarfs \citep{Kawaler2004, Kawaler2014, Ferrario2015}. It is estimated that $10\%$-$20\%$ of all white dwarfs have a magnetic field stronger than 0.1~T, according to the $\sim 250$ known magnetic white dwarfs \citep{Sion2014,Ferrario2015}.

In this section, we describe the test we conducted of the material-shielding mechanism in white dwarfs with known rotational periods and magnetic fields. 
We tested whether the unpolluted DA/DB white dwarfs\footnote{DA and DB white dwarfs refer to those only containing hydrogen and helium, respectively.} and the potentially polluted DBZ/DAZ/DZ white dwarfs\footnote{White dwarfs that are polluted by heavy elements are classified as DBZ, DAZ, and DZ types, some of which were discovered to be volatile-rich.} are subject to our volatile shielded mechanism or not. Due to the limited availability of data pertaining to white dwarf rotations and magnetic fields, the test sample was restricted to a total of nine white dwarfs, as listed in Table \ref{tab1}. With more data on white dwarf rotations and magnetic fields in the future, our model promises more precise testing, enhancing our understanding of the volatile shielded mechanism's applicability to both unpolluted and polluted white dwarfs. We also implemented a test of the refractory-shielding mechanism (where the dust sublimation radius is larger than the corotation radius) on DA, DB, DAZ, DBZ, and DZ white dwarfs (see Appendix \ref{AppE}).

\begin{figure}
    \centering
    \includegraphics[width=0.4\textwidth]{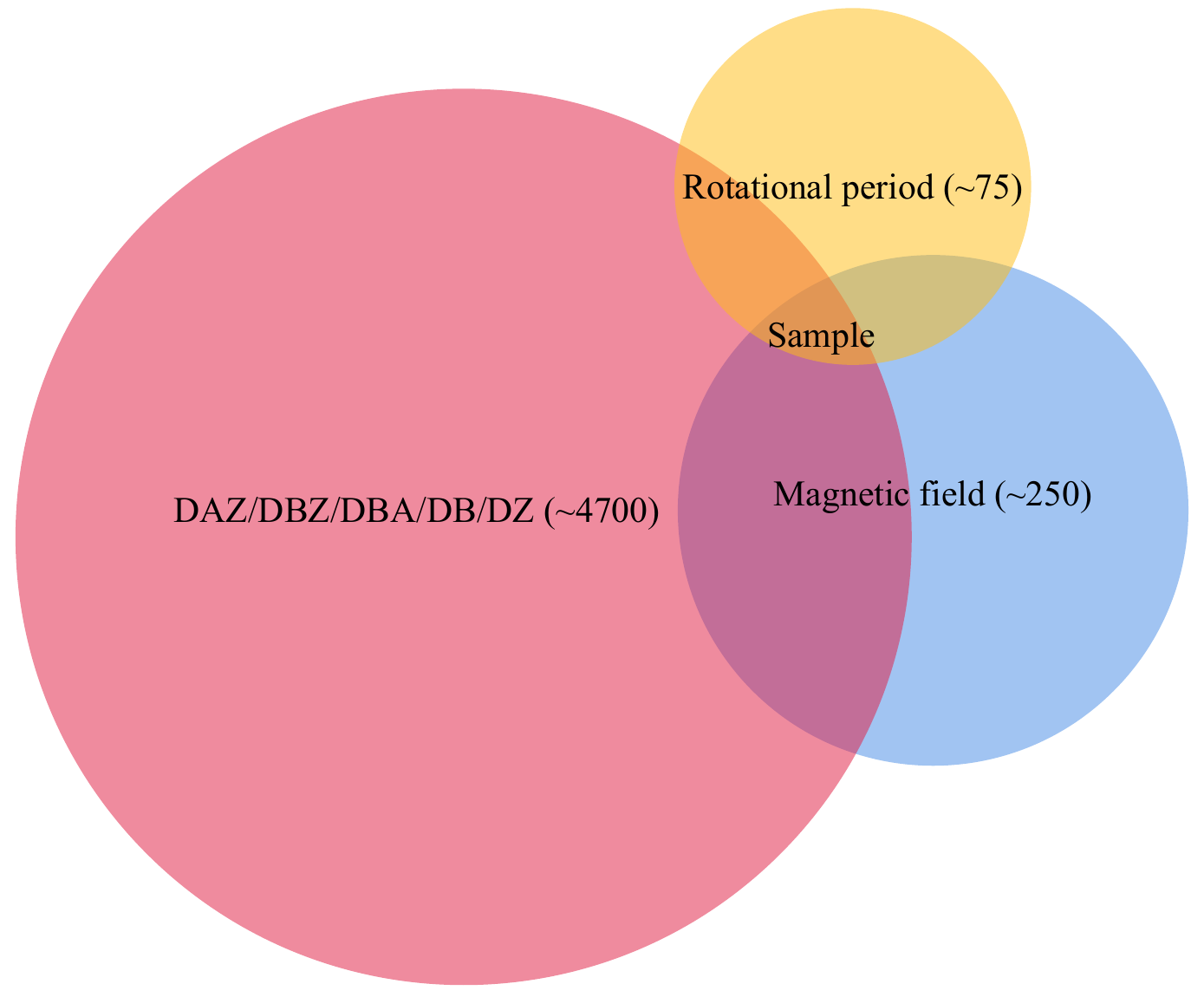}
    \caption{Sampling of our observational tests. Our model includes the composition, surface magnetic field, and spin period of white dwarfs. The relative size of a circle shows the sample size (not to scale) and the numbers in parentheses are the total number of white dwarfs with available data;  there are nine white dwarfs that fulfill our criteria.}
    \label{fig:sample}
\end{figure}

Volatile pollution is typically diagnosed by an excess of carbon, nitrogen, oxygen, or sulfur in the white dwarf's atmosphere. Among the polluted DAZ, DBZ, and DZ white dwarfs, we need further information about whether the heavy elements are volatile-rich or not. In addition, DBA white dwarfs, which contain hydrogen in the helium-dominated atmosphere, are considered to have been potentially water-polluted in the past, as hydrogen does not sink, but accumulates in the atmosphere. On the other hand, in the hydrogen-dominated (DA) white dwarfs, the heavy elements sink quickly, so that the memory of the distant past is lost and we have no clues regarding the pollution. Therefore, we consider DA white dwarfs to be bad test targets whose lack of pollution does not necessarily indicate a shielding mechanism. The sampling strategy is also illustrated in Fig.~\ref{fig:sample}.

\subsubsection{WD2326+049}

WD2326+049 (G29-38) is a DAZ white dwarf that hosts a dust disk at a distance of approximately $1.2~R_\odot$ from the central star \citep{Ballering2022}. The mineral composition of the disk has been compared with the photospheric composition, revealing an excess of volatile elements, namely carbon and sulfur, in the disk \citep[see Figure 14 in][]{Xu2014}. It is important to note that the observed excess of these elements could be partly attributed to the assumptions made in the modeling process, including the assumption of an optically thin disk and the inclusion of niningerite \citep{Reach2005, Reach2009, Xu2014}. Nevertheless, our proposed volatile shielding model offers a plausible explanation for the observed excess of carbon and sulfur.

The white dwarf WD2326+049 (G29-38) possesses a weak magnetic field of $0.07\pm 0.05$T, as reported in a study by \citet{Farihi2018}. However, the measurements of its rotational period have yielded conflicting results, with values of 0.014 days \citep{Koester1998}, 0.042 days \citep{Berger2005}, and 1.35 days \citep{Kleinman1998} reported using different methods. Evidence supporting a period of 0.77 days has also been presented by \citet{Thompson2010}. Based on the conditions required for the shielding of volatile material, illustrated in Figure \ref{fig:condition}, assuming a rotational period of either 0.014 or 0.042 days would satisfy the necessary conditions for shielding if the accretion rate were $10^6~\rm kg/s$. This suggests that volatile material could be shielded from the corotation radius of the white dwarf. Given that the corotation radius for a period of 0.042 days is approximately 0.4~$R_\odot$, which lies between the silicate sublimation radius ($\sim 0.2R_\odot$) and the location of the narrow disk ($1.2R_\odot$ with a width of $0.12R_\odot$), silicate may still be able to pass through the corotation radius in solid form and ultimately be accreted. The situation differs for carbonate and sulfide, as their sublimation temperatures of $\sim$1000K correspond to a sublimation radius of $\sim 0.45 R_\odot$, which is larger than the corotation radius. Consequently, carbonate and sulfide vapor is shielded from the corotation radius by the magnetic field. Outward-moving carbonate gas can effectively recondense, as shown by \citet{Okuya2023}, which leads to the non-detection of gas. As a result, the abundance of carbon and sulfur in the dusty disk exceeds that of the white dwarf atmosphere. A comprehensive investigation of the disk model falls outside the scope of this study, and further scrutiny of this system is deferred to future research endeavors.

\subsubsection{WD0322-019}

WD 0322-019 is classified as a very cool white dwarf and its accretion rate is estimated to be $3.7 \times 10^4 ~ \rm kg/s$ \citep{Farihi2011}. Based on the position of its ice line, which is located at $\sim 20 R_{\odot}$, it is expected that water gas is not shielded in this system, as the ice line is smaller than the corotation radius of the white dwarf, estimated to be $\sim 30 R_\odot$. Within the corotation radius, the magnetic field funnels volatile gases into the white dwarf via sublimation. Hence, WD 0322-019 is expected to be enriched with volatile elements; however,  no such detection has been made to date, including oxygen, which typically exists in rocky materials. The low temperature of the white dwarf makes such detections challenging \citet{Farihi2011}.

\subsubsection{WD1748+708, WD1820+609, WD1829+547, and WD2010+310}

The white dwarfs 1748+708, 1820+609, 1829+547, and 2010+310, all of which possess magnetic fields, exhibit rotational periods that exceed 100 years. As a result, their corotation radii are larger than their ice lines, thereby indicating that the volatile shielding mechanism is inoperative in these stars. Notably, the white dwarfs 1820+609 and 1829+547 are DBA white dwarfs, which is in agreement with our predictions, due to the presence of hydrogen in their atmospheres. On the other hand, the absence of hydrogen in the DB white dwarfs 1748+708 and 2010+310 could be due to the absence of planetary systems or could be a result of an as-yet-unknown mechanism.

\begin{figure}
    \centering
    \includegraphics[width = 0.5 \textwidth]{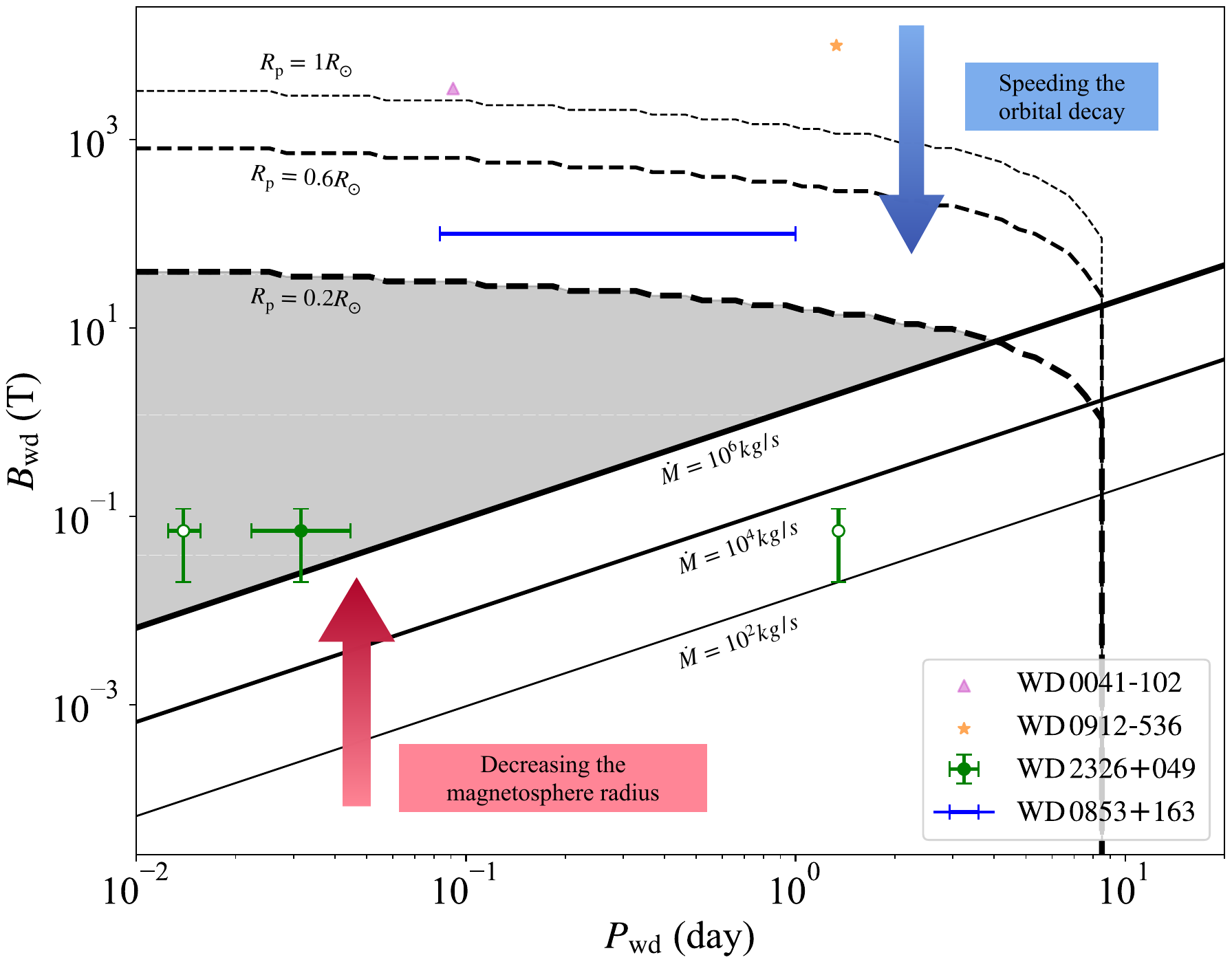}
    \caption{Volatile shielding conditions in the $B_{\rm wd}$--$P_{\rm wd}$ parameter space. The shaded area between the solid and dashed lines indicates the zone where the shield mechanism is in effect. The black solid lines represent the magnetosphere condition, above which $R_{\rm mag}$ is larger than $R_{\rm cor}$, ensuring that gas is shielded from the magnetosphere of the white dwarf.
    The black dashed lines represent the dry-out condition, below which the icy fragments of a comet can be totally dried out before they cross the corotation radius $R_{\rm cor}$ in the process of orbital circularization. The shaded region could expand according to different assumptions on the accretion rate and the pericenters of fragments. The dashed lines are based on the Alfv{\'e}n-wave drag-induced orbital circularization, so in other scenarios of circularization, these dashed line are not valid. The shaded region should also depend on the material properties of planetary pollutants, so there is a large uncertainty on the region above the solid lines.  DBA white dwarfs WD0041-102 and WD0853+163 are out of the shielding regime, consistent with the detection of hydrogen in their atmospheres.  DB white dwarf WD0912-536 is also open to volatiles, but no hydrogen is detected. This might be the result of no planetary system or some unknown mechanism. The blue dotted line denotes the lower limit of the observation on the surface magnetic field of white dwarfs. WD2326+049 has several measurements in the rotational period, two of which are in the volatile-shielding region.}
    \label{fig:condition}
\end{figure}

\subsubsection{WD0041-102, WD0853+163, and WD0912+536}

The magnetic white dwarfs 0041-102, 0853+163, and 0912+536 have a relatively fast rotation rate, resulting in a corotation radius that falls within the ice line. As depicted in Figure \ref{fig:condition}, these white dwarfs could be open to the volatiles in the comet accretion scenario, provided the tidal fragments have a close pericenter. Although the magnetic fields of these white dwarfs are sufficiently strong to hold a magnetosphere beyond the corotation radius, this also means that their orbits rapidly drift toward the corotation radius, thus not enabling them to eliminate the volatile. Since the sublimated volatiles inside the corotation radius are drawn toward the white dwarf surface due to the magnetic field's influence, the shielding mechanism becomes ineffective for them. This observation is consistent with the presence of hydrogen on the surfaces of white dwarfs 0041-102 and 0853+163, both of which are DBA white dwarfs. However, for white dwarf 0912+536, hydrogen is not detected on its surface while it is situated beyond the shielding region. This may be due to the absence of a planetary system or some other unexplained mechanisms.





\section{Summary and conclusion}
White dwarfs have been observed to have atmospheres polluted by heavy elements, which are thought to be accreted from planetary objects \citep{Zuckerman2003}. Refractory pollution, such as silicate and iron elements, is more commonly observed than volatile pollution, which includes carbon, nitrogen, oxygen, and sulfur \citep{Xu2017}. It is currently unclear whether this scarcity of volatile pollution is due to the prevalence of dry pollutants, such as asteroids and rocky planets, or some unknown mechanism that shields the volatiles.

Recently, \citet{Zhang2021} proposed a scenario in which the orbital circularization of long-period comets is achieved through Alfv{\'e}n wing drag, which would allow comets from the exo-Oort Cloud to be accreted onto the white dwarf. However, this scenario does not explain the scarcity of volatile pollution. In light of this problem, we investigate the evolution of volatile content in comets

We find that the volatile content in the tidal fragments of comets can effectively sublimate during the process of orbital circularization. Following sublimation, the resulting volatile vapor may be shielded by the magnetosphere of white dwarfs, provided that the magnetosphere radius is greater than the corotation radius. The effectiveness of this volatile-shielded mechanism is determined by the extent of sublimation occurring outside the corotation radius. The main processes in this scenario are the following:
\begin{description}
\item[(1) Tidal fragmentation and orbital circularization of comets.] Comets are tidally disrupted within the tidal disruption radius. The generated fragments with a maximum size of $\sim 200$~m migrate inward due to the Alfv{\'e}n wing drag and/or dust drag. In this work, we incorporate the Alfv{\'e}n wing drag to model the behavior of exo-Oort Cloud objects.
\item[(2) Volatile sublimation.] After crossing the ice line, the interior of fragments gets heated up to the sublimation temperature of volatile materials, resulting in gas release from the surface. We adopt the physical properties of Solar System comets to study the thermal evolution and vapor evolution. 
\item[(3) Magnetic shielding.] Given that the corotation radius is smaller than the magnetosphere radius, the photoionized vapor is shielded from the corotation radius.
\end{description}

Our findings reveal that objects with a size smaller than 500 m can experience complete volatile loss within 1 Myr. The effectiveness of the volatile-shielded mechanism hinges on whether the orbital circularization timescale is shorter or longer than the dry-out timescale. If it is longer, all the volatile content can be shielded from the corotation radius of the white dwarf. However, for a shorter circularization timescale, smaller icy objects can be completely dried out while larger objects retain some volatile content, leading to partial dryness and the presence of volatile material within the corotation radius. Consequently, the volatile-shielded mechanism cannot provide effective protection in such cases. A surface magnetic field weaker than 100~T could allow a sufficiently long circularization timescale for the volatile-shielded mechanism to work. As we mentioned, the prerequisite of this mechanism is that the magnetosphere radius is larger than the corotation radius. For example, for a white dwarf with a rotational period of two days, the magnetic field must be larger than 1~T.  

By introducing the magnetic shielding of volatile, we extrapolated a correlation between a white dwarf's magnetic field, spin period,  and the composition of the pollutants. We applied our model to nine white dwarfs with known magnetic fields, rotational periods, and atmosphere compositions, and observed that the polluted white dwarf G29-38 (WD2326+049) is volatile shielded in our model, potentially explaining the excess of volatile elements such as C and S in the disk relative to the white dwarf atmosphere. The other eight white dwarfs under investigation {{exhibit receptivity to volatiles.}} This is consistent with the presence of hydrogen in the four investigated DBA white dwarfs. 
However, the lack of hydrogen in the four examined DB white dwarfs cannot be explained by our shielding mechanism and may be attributed to the absence of planetary systems or other unknown mechanisms. We suggest that future observations of white dwarfs with different magnetic fields and rotational periods will provide more insights into the efficiency of the volatile-shielded mechanism. Additionally, our model is highly sensitive to the orbital evolution and material properties of exocomets, and a comparison between observations and our model may lead to valuable constraints on the properties and origins of pollutants.

\begin{acknowledgements}
 We thank Yun Zhang and Siyi Xu for useful discussions. Wen-Han Zhou would like to acknowledge the funding support from the Chinese Scholarship Council (No.\ 202110320014). This work is partly supported by the National Natural Science Foundation of China under grant No. 11903089, the Guangdong Basic and Applied Basic Research Foundation under grant Nos. 2021B1515020090 and 2019B030302001, the Fundamental Research Funds for the Central Universities, Sun Yat-sen University under grant No. 22lgqb33, and the China Manned Space Project under grant Nos. CMS-CSST-2021-A11 and CMS-CSST-2021-B09.
\end{acknowledgements}

\appendix

\section{Tidal disruption of comets}
\label{AppA}
\subsection{Tidal disruption radius}
A comet  can break up due to the tidal force once its orbit is within the tidal disruption radius $R_{\rm tidal}$. For a gravity-dominated object (>200~m), the tidal disruption radius is \citep{Sridhar92}
\begin{equation}
    R_{\rm tid,grav} = 1.05 \left( \frac{M_\star}{\rho} \right)^{1/3} = 1.6 \left(\frac{\rho}{1\rm~g/cm^3} \right)^{-1/3} R_\odot.
\end{equation}
For a strength-dominated object (< a few hundred meters), the tidal disruption radius is \citep{Holsapple07,Holsapple08, Zhang20}
\begin{equation}
    R_{\rm tid,str} = \left( \frac{\sqrt{3}}{4\pi} \right)^{1/3} \left( \frac{5k}{4\pi G r_0^2 \rho^2} + s \right)^{-1/3} \left( \frac{M_\star}{\rho}\right)^{1/3},
\end{equation}
where 
\begin{align}
    & k = \frac{6C \cos \phi}{\sqrt{3} (3-\sin \phi)}  ,\\
    & s = \frac{2 \sin \phi}{\sqrt{3}(3-\sin \phi)}.
\end{align}
Here $r_0$ is the radius of the comet. The cohesive strength $C$ can range from 0.1~Pa to 10~MPa, which introduces uncertainty, and the frictional angle $\phi$ ranges from $25^\circ$ to $50^\circ$. For simplicity, we assume the size of the parent body is larger than 1~km, as the observed abundance of heavy elements in a white dwarf implies an average 100~km sized pollutant (equivalent to the accretion rate of $10^8$~g/s) \citep{Wyatt2014}. Therefore, the comet considered in this work is in the gravity-dominated regime and the corresponding tidal radius is 
\begin{equation}
    R_{\rm tid} \simeq 1.27 R_\odot,
\end{equation}
with $\rho = 2~\rm g/cm^3$.

\subsection{Size distribution}
The fragments of a tidally disrupted comet typically follow a size distribution with a maximum radius $r_{\rm frag,max}$ determined by 
\begin{equation}
    r_{\rm frag,max} = \left( \frac{5k}{4\pi G \rho^2 ({\sqrt{3}M_\star}/{4\pi \rho R^3} - s )} \right)^{1/2}.
\end{equation}
For the region $R < (\sqrt{3}M_\star/4\pi\rho s)^{1/3} \approx R_{tidal,grav}$, $R_{\rm frag,max}$ is approximately
\begin{equation}
    r_{\rm frag,max} = \left( \frac{5k  R^3}{ \sqrt{3} G \rho M_\star} \right)^{1/2} \approx 212 \left( \frac{C}{1~\rm Pa} \right)^{1/2} \left( \frac{R}{1.27 R_\odot} \right)^{3/2} m,
\end{equation}
which is consistent with the lower bound of the size of a rubble pile object \citep{Walsh2018}. The fragments follow a size distribution described by a power law
\begin{align}
    & \frac{{\rm d}N_{\rm frag}}{{\rm d}r_{\rm frag}} = C_N r_{\rm frag}^{-\alpha} ,\\
    & \frac{{\rm d}m_{\rm frag}}{{\rm d}r_{\rm frag}} = \frac{4\pi}{3} C_N \rho r_{\rm frag}^{3-\alpha} 
\end{align}
for $r < r_{\rm frag,max}$, where $\alpha \approx 3.5$ \citep{Wyatt2011} and $C_N$ is constrained by mass conservation:
\begin{equation}
    C_N = (4 - \alpha) r_0^3 r_{\rm frag,max}^{\alpha - 4}.
\end{equation}
The fragments smaller than $r_{\rm frag,min}$ are blown out by radiation pressure \citep{Burns1979, Brouwers2022a}, which is given by
\begin{equation}
\begin{aligned}
     r_{\rm frag,min} = 0.15 \left( \frac{a_0}{\rm au} \right)\left( \frac{R}{\rm R_\odot} \right)^{-1} \left( \frac{L_\star}{0.001 L_\odot} \right)  \left( \frac{M_\star}{0.6 M_\odot} \right)^{-1} \rm \mu m
\end{aligned}
.\end{equation}

\begin{figure}
    \centering
    \includegraphics[width = 0.5\textwidth]{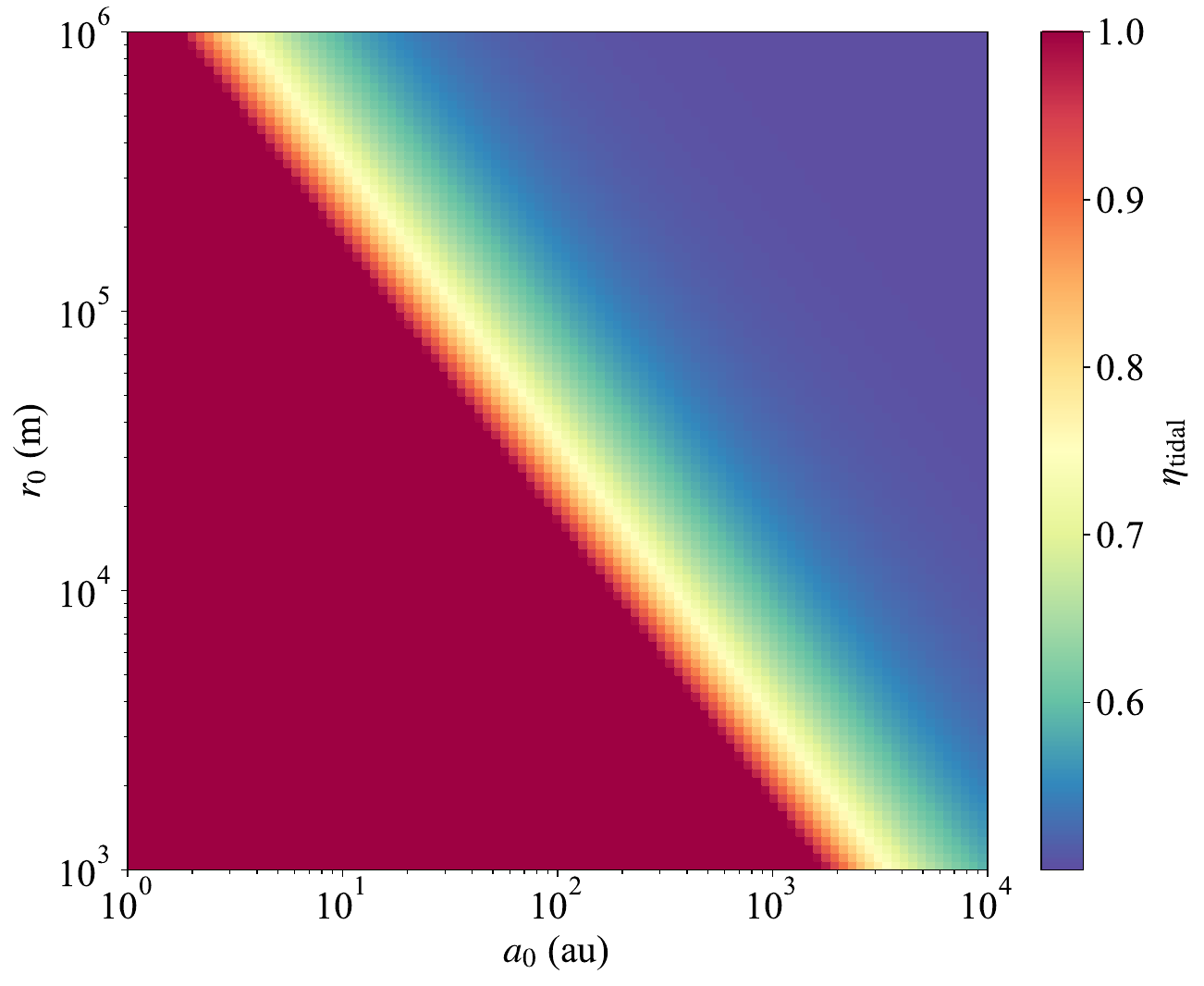}
    \caption{Bound mass fraction (denoted by the color scale) after tidal disruption as a function of the size and the initial semimajor axis of the parent body.}
    \label{fig:tidal_loss}
\end{figure}

\subsection{Orbital distribution of fragments}
After disruption, fragments of the parent body obtain new semimajor axes \citep{Brouwers2022a}
\begin{equation}
\label{eq:a_frag0}
    a_{\rm frag,0} = a_0 \left(1 - \frac{ 2a_0 x r_0}{R_{\rm tid} (R_{\rm tid} + x r_0)} \right)^{-1},
\end{equation}
and new eccentricities
\begin{equation}
    e_{\rm frag,0} = 1 - \frac{R_{\rm tid} + x r_0}{a_{\rm frag,0}},
\end{equation}
where $a_0$ and $r_0$ are the semimajor axis and radius of the parent body, respectively. The parameter $x$ in the range of (-1,1) describes the initial location of the fragment before the disruption. The inner bound of the fragments is expressed as 
\begin{equation}
    a_{in} = a_0 \left(1 +\frac{ 2a_0  r_0}{R_{\rm tid}(R_{\rm tid} +  r_0)} \right)^{-1}.
\end{equation}
Fragments could have a very far outer bound and even be unbound if the parent body size is larger than a critical size \citep{Malamud2020}
\begin{equation}
    r_0 > \frac{r_R^2}{2a_0 - r_R} \approx \frac{r_R^2}{2a_0} = 187.5 \left( \frac{r_R}{1 R_\odot}\right)^2 \left( \frac{a_0}{10 \rm~au} \right) \rm km,
\end{equation}
which is obtained by setting $a_{\rm frag,0}<0$ in Eq. (\ref{eq:a_frag0}). In this case, the remaining mass ratio due to tidal disruption is approximately
\begin{equation}
    \eta_{tidal} = \frac{1}{2} + \frac{r^2}{2r_0^2 (2 a_0 + r)}.
\end{equation}
Figure \ref{fig:tidal_loss} shows the remaining mass ratio with dependence on the size and the semimajor axis of the parent body. It can be seen that objects larger than 1 km from the Oort Cloud ($\sim 10^4\,$au) lose about half of their mass, and those from the Kuiper belt $(\sim 50 \rm~ au)$ could maintain most of their mass if they are smaller than 100 km.

\section{Circularization by Alfv{\'e}n drag}
\label{AppB}
When fragments move through a magnetized flowing plasma, Alfvén wave is induced and dissipates the orbital energy with a power \citep{Drell65, Neubauer80}
\begin{equation}
    P \simeq \frac{\pi U^2}{\mu_0 \bar v_A},
\end{equation}
where $U = r_{\rm frag}v_{\rm rela}B$ is the electric potential of the induced electric field across the radius $r_{\rm frag}$, $v_{\rm rela}$ is the relative speed of the fragment to the magnetic field, $B$ is the local magnetic field flux density, and $\mu_0 = 4\pi \times 10^{-7} \rm N~A^{-2}$ is the permeability of free space. We neglect the spin of the white dwarf such that $v_{\rm rela}$ is equivalent to the orbital speed $v$. The parameter $\bar v_A$ is estimated as $3 \times 10^7~\rm m/s$ in the low gas number density regime \citep{Zhang2021}.

The dissipated energy within an orbital period is
\begin{equation}
\label{eq:Delta_E1}
    \Delta E = - \frac{P}{m} \Delta t_{\rm eff},
\end{equation}
where $m$ is the fragment mass and $\Delta t_{\rm eff}$ is the effective time of energy dissipation. The local magnetic field $B$ decreases rapidly with the distance to the white dwarf as 
\begin{equation}
    B = B_\star \left( \frac{R}{R_\star}\right)^3.
\end{equation}
Therefore, the orbital energy dissipation is efficient only around the pericenter $R_{\rm p} = a(1-e)$ when the orbital is highly eccentric. The effective time $\Delta t_{\rm eff}$ can be expressed as
\begin{equation}
\label{eq:Delta_t_eff}
    \Delta t_{\rm eff} = \frac{a}{\Omega_{\rm peri}}
,\end{equation}
where $\Omega_{\rm peri}$ is the angular speed at the pericenter
\begin{equation}
\label{eq:Omega_peri}
    \Omega_{\rm peri} = \frac{\sqrt{GM_\star R_{\rm p}(1+e)}}{R_{\rm p}^2}
\end{equation}
and $\alpha \simeq \pi/2.24$ is an empirical value \citep{Zhang2021}. 
Combining above equations, we obtain the energy change
\begin{equation}
    \Delta E \simeq - \frac{3\alpha B_\star^2 R_\star^6}{2 \rho r_0 \mu_0 \bar v_A p^5} \sqrt{G M_\star \over p}.
\end{equation}
Since the aphelion of fragments is distant where the magnetic effect is inefficient, the distance of the pericenter to the white dwarf $R_{\rm p}$ does not change a great deal. Therefore, the energy change is nearly constant over the orbital evolution.


\begin{figure}
    \centering
    \includegraphics[width=0.5\textwidth]{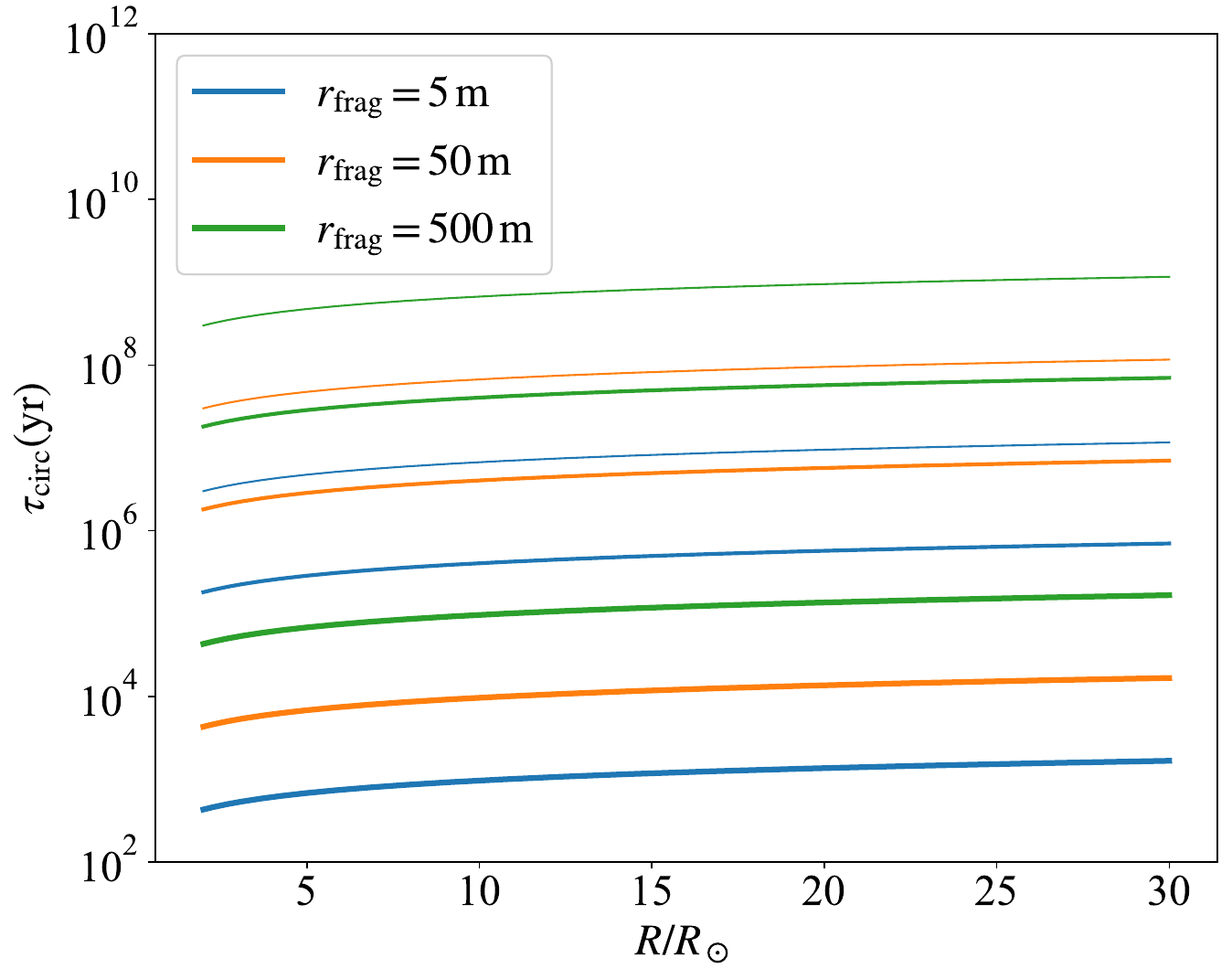}
    \caption{Timescale for the fragments of different sizes to be circularized by the Alfv{\'e}n-wave drag as a function of the distance to the white dwarf, accounting for pericenter distances of $0.2 R_\odot$, $0.6 R_\odot$, $1 R_\odot$ denoted by different line widths from the thickest to the thinnest.}
    \label{fig:circularization_timescale}
\end{figure}

The change of the semimajor axis and the eccentricity in an orbital period due to the energy dissipation are
\begin{equation}
    \Delta a  = -\frac{a \Delta E }{E}
\end{equation}
and
\begin{equation}
    \Delta e = -(1-e) \frac{\Delta E}{E}, 
\end{equation}
respectively. With the orbital energy $E = -G M_\star / 2a$ and the orbital period $P_{frag} = 2\pi \sqrt{a^3/GM_\star}$, we obtain the orbital migration rate
\begin{equation}
    \dot a \simeq {\Delta a \over P_{frag}} = - { \Delta E \over \pi \sqrt{GM_\star}} \sqrt{a},
\end{equation}
which gives
\begin{equation}
\label{eq:a}
    a \simeq \left(\sqrt{a_0} + {\Delta E \over 2\pi \sqrt{GM_\star}}t\right)^2,
\end{equation}
and 
\begin{equation}
    e \simeq 1 - p \left(\sqrt{a_0} + {\Delta E \over 2\pi \sqrt{GM_\star}}t\right)^{-2}.
\end{equation}
Figure \ref{fig:circularization_timescale} shows the circularization timescale of fragments of different sizes caused by the magnetic drag. The timescale of this orbital circularization is
\begin{equation}
\begin{aligned}
\label{eq:tau_B,circ}
    \tau_{B,circ} &= {2\pi \sqrt{GM_\star a_0 } \over -\Delta E}    \\
    & = 2.7 \left(\frac{r}{1~\rm m} \right)\left(\frac{B_\star}{100~\rm T} \right)^{-2} \left(\frac{a_0}{100~\rm au} \right)^{1/2}  \left(\frac{q_0}{0.2 R_\odot} \right)^{11/2}  {~\rm kyr}.
\end{aligned}
\end{equation}Here we use the approximation $\dot a \simeq \Delta a/P_{\rm frag}$, which is valid when $t \gg P_{\rm frag}$. A more precise description of the orbital evolution is given as a function of the number of orbits \citep{Zhang2021}.

\section{Photoionization of hydrogen}
\label{AppC}
The volatile vapor is ionized by stellar radiation on a timescale
\begin{equation}
    \tau_{ion} = 1/P_{ion}.
\end{equation}
Here $P_{ion}$ is the probability per unit time of photoionization, given by
\begin{equation}
    P_{ion} = \int_{\nu_L}^\infty {L_\nu(T) \sigma_H(\nu) \over 4 \pi r^2 h\nu}  d\nu
,\end{equation}
where $L_\nu(T)$ is the specific luminosity, $\sigma_H$ is the photoionization cross section, and $h\nu_L = 13.6\rm~eV$ with $h$ denoting the Plank constant. The specific luminosity $L_\nu$ is obtained by
\begin{equation}
    L_\nu (T) = 4 \pi^2 R_{\rm wd}^2 B_{\nu} (T),
\end{equation}
where $B_\nu$ describes the blackbody radiation
\begin{equation}
    B_\nu (T ) = {2 \nu^2 \over c^2} {h \nu \over \exp ({h \nu /kT_\star}) - 1}.
\end{equation}
The photoionization cross section $\sigma_H$ approximately follows a power law
\begin{equation}
    \sigma_H \approx \sigma_0 \left( \frac{\nu}{\nu_L}   \right)^{-3} 
\end{equation}
for photons with frequency $\nu > \nu_L$.
Here $\sigma_0 = 6.3 \times 10^{-22} \rm~m^2$. The ionization timescale is 
\begin{equation}
    \tau_{ion} = 1.5 \times 10^{-5}   \left(\frac{R}{R_\odot}\right)^{2}    \rm yr.
\end{equation}

\section{Photoionization equilibrium}
\label{AppD}
The photoionization equilibrium is obtained when the photoionization rate is equal to the recombination rate. The timescale of radiative recombination is
\begin{equation}
    \tau_{rec} = \frac{1}{n_e \alpha_A},
\end{equation}
where $n_e$ is the number density of electrons, and $\alpha_A$ is the radiative recombination coefficient that can be estimated by
\begin{equation}
    \alpha_A \approx 4.13 \times 10^{-13} \left( \frac{T_\star}{10^4 \rm K} \right)^{-0.71 - 0.011\ln(T_\star/10^4 \rm K)}
\end{equation}
for an optically thin regime. The electron number density can be assumed as $n_e = n_H = 2n_{\rm H_2O}$, where
\begin{equation}
\label{eq:n_H2O}
    n_{\rm H_2O} = \frac{\dot M}{4\pi r^2 v_g m_{\rm H_2O}},
\end{equation}
where $\dot M$ is the mass accretion rate, $m_{\rm H_2O}$ is the mass of $\rm H_2O$ molecule, and the gas incoming velocity $v_g$ is assumed to be less than the Kepler velocity 
\begin{equation}
\label{eq:v_g}
    v_g \leq \sqrt{\frac{G M_{\rm wd}}{r}}.
\end{equation}
Therefore, the recombination timescale is 
\begin{equation}
    \tau_{rec} = 1.8 \times 10^4   \left(\frac{R}{R_\odot}\right)^{3/2}    \rm yr \gg \tau_{ion}.
\end{equation}
At the photoionization equilibrium
\begin{equation}
    \frac{n_{HI}}{\tau_{ion}} = \frac{n_p}{\tau_{rec}},
\end{equation}
the ionization ratio is 
\begin{equation}
    \eta_{ion} = \frac{n_p}{n_{HI}+n_p} = \frac{\tau_{rec}}{\tau_{rec}+\tau_{ion}} \approx 1,
\end{equation}
which indicates that near the white dwarf, gas is nearly completely ionized on a timescale much shorter than the orbital {circularization} timescale.


\section{Test on refractory-shielding mechanism}
\label{AppE}

\begin{figure}
    \centering
    \includegraphics[width = 0.5\textwidth]{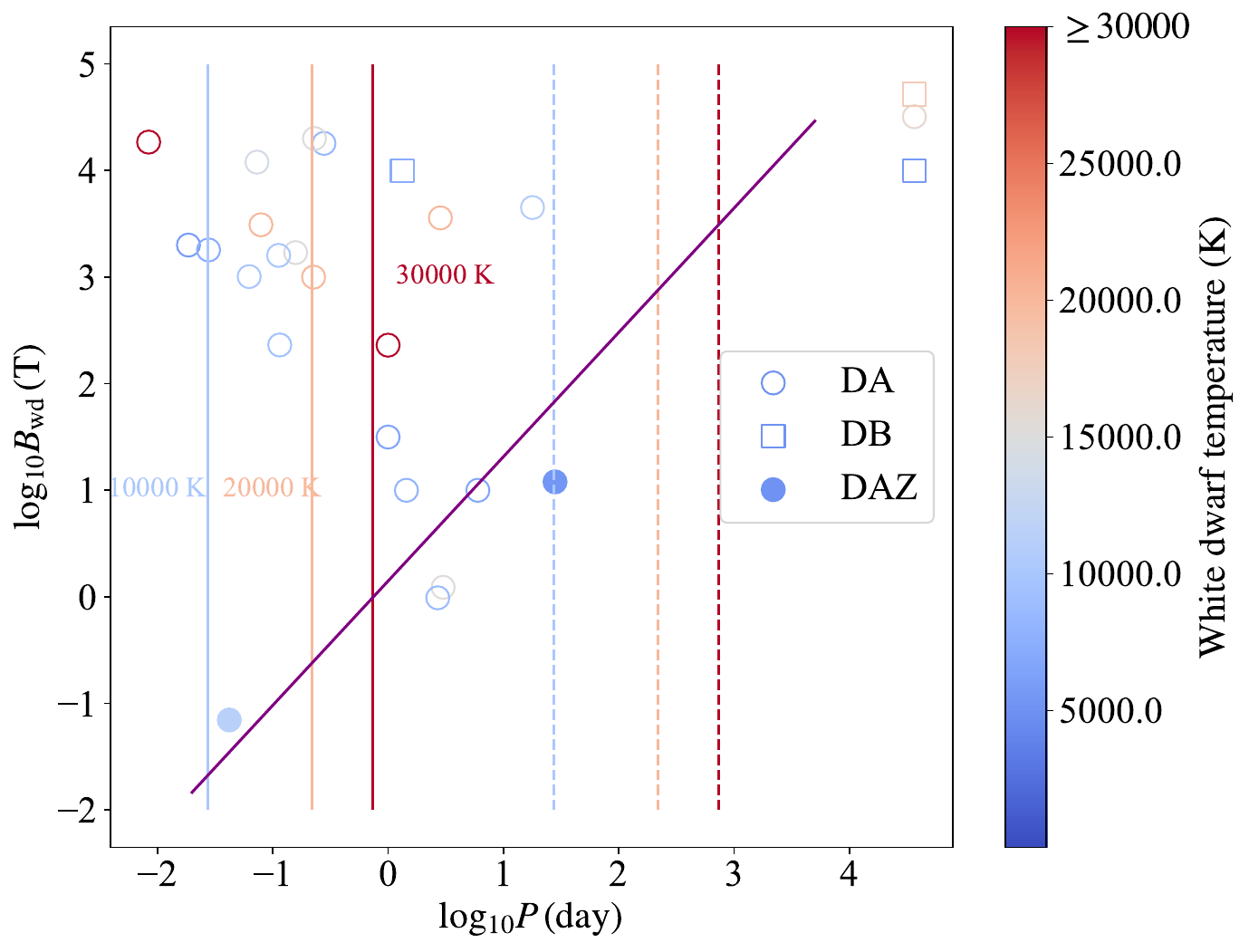}
    \caption{Material shielding conditions in the $B_{\rm wd}$–$P_{\rm wd}$ parameter space. DA, DB, and DAZ white dwarfs are denoted as open circles, open circles, and solid circles, respectively. The purple line represents the magnetosphere condition, above which the magnetic effect is active. The vertical lines are the rotational periods where the corotation radius equals the sublimation radius for refractory materials (solid lines) and volatile materials (dashed lines). To the left of the vertical lines, the shielding effect works as the corotation radius is smaller than the sublimation radius. Overall, the material-shielding regime is represented by the upper left region of the plot separated by lines.}
    \label{fig:unpolluted}
\end{figure}

White dwarfs of spectral types DB, DAZ, DBZ, and DZ are suitable targets for examining the refractory-shielding phenomenon. However, the sample size is currently limited, as only two DAZ and three DB white dwarfs with known magnetic fields and periods are available. To broaden the scope of the study, 21 DA white dwarfs were also included in the sample pool. Nevertheless, it is worth noting that inferring the existence of the refractory-shielding phenomenon in DA white dwarfs from the absence of pollution might not be entirely accurate, given that heavy elements in DA white dwarfs tend to rapidly sink. Figure \ref{fig:unpolluted} presents the result, which does not reveal any discernible feature. Therefore, we recommend that further investigations be conducted in the future to study this issue.

\end{document}